\input ./style/arxiv-general.cfg
\documentclass[aoas,MSNbibl,nameyear,seceqn,dvips]{arximspdf}
\makeatletter
   \@ifpackageloaded{graphicx}{}{\usepackage{graphicx}}
\makeatother

\doi{10.1214/15-AOAS853}
\volume{9}
\issue{4}
\pubyear{2015}
\firstpage{1864}
\lastpage{1888}
\docsubty{FLA}


\begin{document}
\begin{frontmatter}

\title{Bayesian analysis of traffic flow on interstate I-55: The LWR model}
\runtitle{Bayesian analysis of traffic flow}

\begin{aug}
\author[A]{\fnms{Nicholas}~\snm{Polson}\corref{}\ead
[label=e1]{ngp@chicagobooth.edu}}
\and
\author[B]{\fnms{Vadim}~\snm{Sokolov}\ead[label=e2]{vs@anl.gov}}
\runauthor{N. Polson and V. Sokolov}
\affiliation{University of Chicago and Argonne National Laboratory}
\address[A]{Booth School of Business\\
University of Chicago\\
Chicago, Illinois 60637\\
USA\\
\printead{e1}}
\address[B]{Argonne National Laboratory\\
Lemont, Illinois 60439\\
USA\\
\printead{e2}}
\end{aug}

%
\received{\smonth{2} \syear{2015}}
%
\revised{\smonth{9} \syear{2015}}

%
\begin{abstract}
Transportation departments take actions to manage traffic flow and
reduce travel times based on estimated current and
projected traffic conditions. Travel time estimates and forecasts
require information on traffic density
which are combined with a model to project traffic flow such as the
Lighthill--Whitham--Richards (LWR) model. We develop a particle
filtering and \mbox{learning} algorithm to estimate the current traffic
density state and the LWR parameters. These inputs are related to the
so-called fundamental diagram, which describes the relationship between
traffic flow and density. We build on existing methodology by allowing
real-time updating of the posterior uncertainty for the critical
density and capacity parameters. Our methodology is applied to traffic
flow data from interstate highway I-55 in Chicago. We provide a
real-time data analysis of how to learn the drop in capacity as a
result of a major traffic accident. Our algorithm allows us to
accurately assess the uncertainty of the current traffic state at shock
waves, where the uncertainty is a mixture distribution. We show that
Bayesian learning can correct the estimation bias that is present in
the model with fixed parameters.
\end{abstract}

%
\end{frontmatter}

\section{Introduction}
Effectively managing traffic flow to reduce congestion can improve
communities by reducing travel times, reducing pollution and improving
economic efficiency. Transportation departments use information on
current and projected travel times to adjust ramp metering and traffic
lights; travelers use projected travel times to make travel plans and
to adjust departure times, transportation mode and route. Estimated
travel times are developed using sophisticated models of traffic flow
that begin with observations on speed and density and develop estimates
of road capacity based on estimates of current density and flow.

In their seminal paper, \citet{lighthill1955kinematic} describe the
theory of kinematic wave motion which they apply to modeling highway
traffic flow. \citet{richards1956shock} independently proposed a similar
application. The key assumption is a relationship between traffic flow
and density. A model is calibrated using the characteristics of road
segments, such as the number of lanes, free-flow speed and road type.
These characteristics themselves do not explain all the variation in
model parameters and estimates need to be assessed using observations
on current speed, density and lane configurations at sparse points
throughout the network. Much of the recent improvement in travel time
estimation and forecasting has come from improving the estimates of
network characteristics [\citet
{dervisoglu2009automatic,muralidharan2009imputation}].

Usually, underlying traffic data is sparse. We observe specific points
in a traffic network using fixed loop-sensors or at random points via
GPS-equipped probe vehicles. Underlying road capacity might vary [\citet
{brilon2005reliability}] as drivers change speed in response to
congestion, weather conditions and the behavior of other drivers, as
well as the number of available lanes change due to weather conditions,
traffic issues and other events. Accurately estimating road capacity
from sparse and noisy observations of traffic speed and density at
points in the traffic network is a significant challenge and improving
on these estimates will lead to better travel time forecasts.

Our approach develops a particle filtering and learning algorithm for
estimating road capacity. We build on existing estimation methods in a
number of ways:
\begin{longlist}[2.]
\item[1.] Incorporation of sequential parameter learning in order to update
the model in real time.
\item[2.] A predictive likelihood particle filter that provides an
efficient estimation strategy and is less sensitive to measurement outliers.
\end{longlist}

We apply our methodology to traffic flow data from Chicago's interstate
I-55 highway and show how parameter learning effectively handles a
dynamic environment, including shock waves. Bayesian learning, which is
central to our methodology, corrects for bias that results from
estimation with fixed parameters. We also show that our algorithm
identifies the drop in road throughput as a result of an accident.

Particle filtering allows for posterior estimation of the most recent
state. The low computational complexity of particle filtering makes
frequent updating feasible, whereas MCMC's computational cost grows
linearly with the length of the data. For previous MCMC applications in
transportation, see \citet{tebaldi1998bayesian} for inferring network
route flows, and \citet{westgate2013travel} for travel time reliability
for ambulances using noisy GPS for both path travel time and individual
road segment travel time distributions. \citet
{anacleto2013multivariate} develop a dynamic Bayesian network to model
external intervention techniques to accommodate situations with
suddenly changing traffic variables. \citet{chiou2013} provide a
nonparametric prediction model for traffic flow trajectories, and \citet
{chiou2012dynamical} proposes using a functional mixture prediction approach.

Previous work on estimating traffic flows use extensions of the Kalman
filter and rely heavily on Gaussianity assumptions; see \citet
{gaz71,sch10,wang05,work08}. \citet{sun2003highway} considered mixture
Kalman filters for traffic state estimation in the context of ramp
metering control. Particle filters have previously been applied to
traffic flow problems; see \citet{mih07} who use the evolution dynamics
as a proposal distribution before resampling, the so-called bootstrap
or sampling/importance resampling (SIR) filter. We improve the
efficiency for inference and prediction with a fully adopted filter and
our approach naturally incorporates particle learning. We build on
existing work on parameter learning in transportation. For example,
\citet{dervisoglu2009automatic} develop a quantile regression
methodology that re-estimates parameters every five minutes based on
traffic flow and density measurements. \citet{wang05} propose an
extended Kalman filter with boundary condition estimation. The
advantage of particle filtering over traditional Kalman filtering is the
ability to handle nonnormal posterior distributions that result, for
example, from nonlinearity. Section~\ref{sec:mixture} shows that the
distribution of uncertainty about state is a mixture at some points in
time and this leads us to use particle filters, that do not rely on
normality assumption.

Real-time estimation and short-run prediction of traffic conditions
play a key role in Intelligent Transportation Systems (ITS). Current
Vehicle Navigation Systems and Traffic Management Systems use forecasts
of traffic flow variables, such as traffic volume, travel speed or
traffic density ranging from 5--30 minutes ahead. There are a number of
real-world applications:
\begin{description}
\item[Advanced Traveler Information Services (ATIS).]
Multiple studies have\break shown the positive impacts of providing
information on traffic flow conditions to the public [\citet
{chorus2006use}], as it can potentially lead to congestion relief
[\citet{arnott1991does}]. Travel information is provided in multiple
ways, for example, by transportation system managers such as local
departments of transportation via variable message signs or radio,
automakers through in-dash navigation, technology companies through
phone apps or web, fleet managers and transit operators.

\item[Transportation Planning.] Benefits of Intelligent Transportation
Systems are studied by local governments based on system performance
data before and after ITS is deployed. An accurate comparison of the
benefits to travel times requires efficient estimation of the network states.

\item[Control of Transportation Operation.]
For traffic control applications, we\break need to efficiently estimate the
formation of traffic congestion. Accurate knowledge of the current
state allows transportation system managers to provide a reasonable
forecast of traffic conditions and to improve traffic flows using such
techniques as ramp metering and speed harmonization.
\end{description}

The rest of the paper proceeds as follows. Section~\ref{sec:lwr}
develops a statistical treatment for the LWR model by representing it
as a nonlinear state-space model. The key input to the LWR model, the
fundamental diagram (or flux function), which links traffic flow and
density is discussed. The parameters of the fundamental diagram need to
be estimated in an online fashion. Section~\ref{sec:analysis} provides
a particle filtering algorithm for inference and prediction that
provides online real-time inference for fundamental diagram parameters
and traffic density state. Section~\ref{sec:study-capacity} illustrates
how our methodology can learn road capacity when applied to data
measured during a major highway accident. Section~\ref{sec:numerical}
illustrates our methodology with a simulation study of rush hour
traffic on Chicago's I-55. Finally, Section~\ref{sec:conclusion}
concludes with directions for future research.

\section{LWR traffic flow model}\label{sec:lwr}

\subsection{Model and data description}
\begin{figure}

\includegraphics{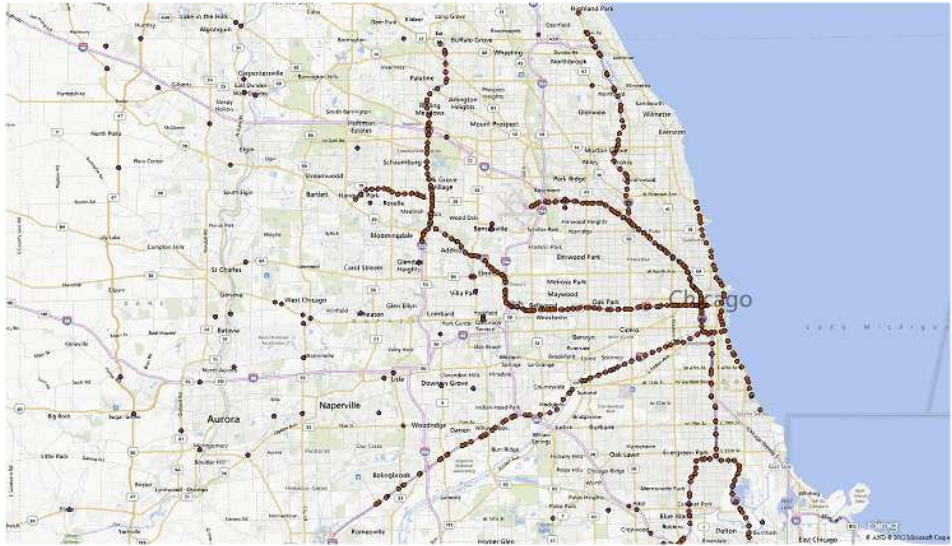}

\caption{Locations of the Loop Detectors in Chicago.}\label{fig:gcm-locations}
\end{figure}

Traffic flow data is available from the Illinois Department of
Transportation; see Lake Michigan Interstate Gateway Alliance
(\surl{http://www.travelmidwest.com/}), formally the Gary--Chicago--Milwaukee
Corridor (GCM). The data is measured by loop-detector sensors installed
on interstate highways. Loop-detector is a simple presence sensor that
measures when a vehicle is present and generates an on/off signal. There
are over 900 loop-detector sensors that cover the Chicago metropolitan
area. Figure~\ref{fig:gcm-locations} illustrates the locations of the
detectors in the region.
Since 2008, Argonne National Laboratory has been archiving traffic flow
data every five minutes from the grid of sensors. Data contains
averaged \textit{speed}, \textit{flow} and \textit{occupancy}.
Occupancy is defined as percent of time a point on the road is occupied
by a vehicle and flow is the number of off-on switches. Illinois uses a
single loop-detector setting and speed is estimated based on the
assumption of an average vehicle length.

%
%

\subsection{Traffic flow parameters}\label{sec:traffic-parameters}
The primary variable of interest is traffic density, which is a
macroscopic characteristic of traffic flow and the control variable of
interest in transportation system management strategies. Traffic
density is defined as a number of vehicles per unit of length.
Densities vary between zero and jam density which corresponds to
vehicles being bumper-to-bumper. Typically, jam density value is around
1 vehicle per 6.5 meters per lane. Another important value related to
density is the critical density, denoted by the density level at which
the maximum flow (throughput) is achieved. The maximum flow measured in
vehicles per unit of time is called capacity, it is typically achieved
at density level of around 1 vehicle per 32 meters.

It is natural to divide the flow regimes roughly into two
subcategories. Density values up to 1 vehicle per 32 meters correspond
to a free-flow regime, when there are no interactions between the
vehicles, and vehicles travel at the desired speed. The second regime
corresponds to densities above 1 vehicle per 32 meters; at roughly this
density vehicles start interacting with each other and that leads to
slow downs and flow reduction.

Our observed data that comes from a presence sensor is occupancy rather
than density. Occupancy is defined as percentage of time a point on a
road segment was occupied by a vehicle, thus it varies between 0 (empty
road) and 100 (complete stand still). Assuming the average vehicle
length does not vary over time, the density and occupancy are related
through a simple linear transformation [\citet{may1990traffic}].
Throughout the paper we assume a constant vehicle length for every
sensor in the region and treat density and occupancy interchangeably.

Two other macroscopic traffic flow parameters, namely. Speed and flow,
are related through the following relation:
%
\begin{equation}
\label{eqn:speed-flow-density} v(x,t) = \frac{q(x,t)}{\rho(x,t)},
\end{equation}
where $v ={}$speed (miles per hour),
$q ={}$flow (vehicles per hour),
$\rho ={}$density (vehicles per lane-mile).

The three traffic flow parameters can change over space and time.
Tracking these flow parameters can be particularly challenging due to
discontinuities in them that are called shock waves. A shock wave can
be a platoon of vehicles moving on an otherwise empty road, thus we
have a nonzero density propagating in time and space. In other cases,
the shock wave corresponds to a change in the flow regime, when
fast-moving vehicles reach the end of a congestion queue and need to
abruptly slow down, or vice versa, when we have queue dissipation and
vehicles leave a bottleneck and can revert to the desired travel speed.

\subsection{The LWR model and fundamental diagram}
In Section~\ref{sec:traffic-parameters} we considered traffic flow as a
function of location $x$ and time $t$. The flow-density relation, which
is called the \textit{fundamental diagram}, allows us to calculate flow
via density $q(x,t) = q(\rho(x,t))$. The LWR model is a macroscopic
traffic flow model. It is a combination of a conservation law defined
via a partial differential equation and a fundamental diagram. The
nonlinear first-order partial differential equation describes the
aggregate behavior of drivers. The density $\rho(x,t)$ and flow
$q(x,t)$, which are continuous scalar functions, satisfy the equation
%
\begin{equation}
\label{eqn:lwr1} \frac{\partial\rho(x,t)}{\partial t} + \frac{\partial
q(x,t)}{\partial x} = 0.
\end{equation}
Derivation of the model is presented in Appendix~\ref{app:model-derivation}.
This equation can be solved numerically by discretizing time and space.
In its simplest form, imagine a homogeneous road segment (with no
change in number of lanes and no intersections) cut into $M$ cells. Let
$\rho_i$ be the density in cell $i$ (in veh${}/{}$m) and $q_i$ the exit flow
of cell $i$ (in veh${}/{}$s). For a road segment, with given boundary
conditions, the LWR computes the conditions inside the domain. Boundary
conditions can be either measured by fixed sensors such as loop
detectors or estimated from GPS probe data based as shown by \citet
{claudel2010lax}. Statistical inference is required to update the
missing states, learn the parameters of interest and predict forward
using the dynamics of the LWR model, based on noise and possibly
partially measured boundary conditions.

%
\begin{figure}[b]

\includegraphics{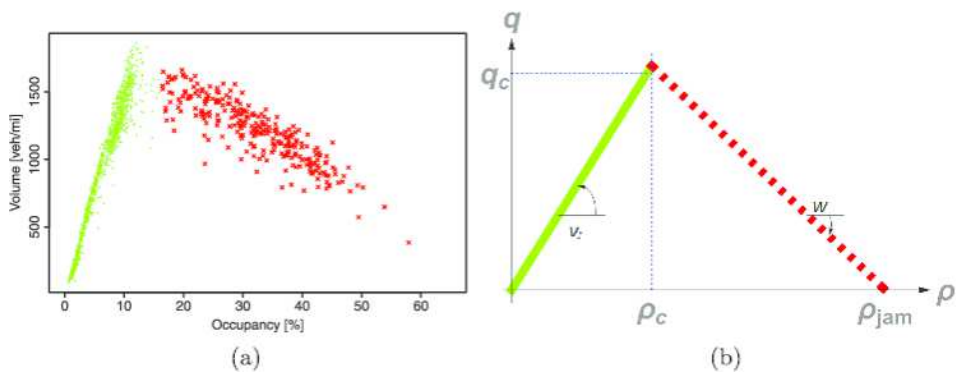}

\caption{Fundamental diagram.
\textup{(a)}~Measured occupancy-flow relation.
\textup{(b)}~Triangular fundamental diagram.
The left panel \textup{(a)} shows the
occupancy-flow relation based on measured data on I-55 North Bound. The
right panel \textup{(b)} shows theoretical shape of the fundamental and the
parameters that describe the diagram. On both panels the {left} part of the diagram [triangles in~\textup{(a)} and solid line
in~\textup{(b)}] describes the density variations for free-flow traffic and the
{right} part [crosses in \textup{(a)} and dashed line in \textup{(b)}] describes
congested traffic.}\label{fig:fd_measured}
\end{figure}

An important feature of the LWR model is the emergence of a shock wave
of traffic due to the density-dependent local propagation velocities.
The fundamental diagram is central to its specification. The diagram
describes a functional relation between flow and density. For example,
Figure~\ref{fig:fd_measured}(a) illustrates empirical data of volume
versus occupancy. The theoretical form of the so-called triangular
fundamental diagram is shown in Figure~\ref{fig:fd_measured}(b). It has
two velocities of density variations: one for free-flow traffic (green)
and one for congested traffic (red). This specification allows for an
efficient Godunov scheme, to solve the nonlinear evolution dynamics. We
need to provide the model with an accurate assessment of the current
density state vector and the parameters of the fundamental diagram. The
\emph{fundamental diagram} is a key input into the specification of the
LWR model, which expresses the relationship between traffic density and
flow. Figure~\ref{fig:fd_measured} motivates the choice of a so-called
triangular diagram by showing the empirical flow and occupancy for a
highway segment in the Chicago metropolitan area.

Throughout our analysis we assume a homogeneous road segment with a
fundamental diagram that does not depend on time and space. By
homogeneous, we mean that the road segment has homogeneous width and
number of lanes and no intersections or traffic merge/diverge sections.

The analytical formula for the triangular fundamental diagram is as follows:
%
\begin{equation}
\label{eqn-newell-fd} q(\rho) = \cases{ \displaystyle\frac{q_c}{\rho
_c}\rho, &\quad$\rho
<\rho_c$,
\cr
\displaystyle q_c \frac{\rho_{\mathrm{jam}}-\rho}{\rho_{\mathrm{jam}} - \rho_c}, &
\quad$\rho\ge\rho_c$,}
\end{equation}
where $q_c ={}$representing the critical flow (capacity),
$\rho_c ={}$critical density,
$\rho_{\mathrm{jam}} ={}$jam density.
We denote the set of three parameters by $\phi= (q_c,\rho_c, \rho_{\mathrm{jam}})$.

The velocity of a shock wave propagation on a road segment can be
calculated using the fundamental diagram parameters, via the
Rankine--Hugoniot relation [\citet{leveque2002finite}]. It determines
the shock wave velocity as the velocity of the shock $w$ times the jump
in density which equals the jump in flow in the two regions separated
by the shock where
%
\begin{equation}
\label{eqn:r-h} w = \frac{q(\rho_l) - q(\rho_r)}{\rho_l - \rho_r} \quad \mbox{and} \quad v_f =
\frac{q_c}{\rho_c}.
\end{equation}
The direction of the shock wave propagation depends on the sign of
$q(\rho_l) - q(\rho_r)$. Here $v_f$ is a free-flow speed on a link and
$q_c = \max_{\rho} q(\rho)$ is the critical flow or capacity of the
link. Correspondingly, $\rho_c = \arg\max_{\rho} q(\rho)$ is called the
critical density. The pair $(q_c,\rho_c)$ is the traffic flow breakdown
point for a road segment.

Calibrating the model parameters can be done in a number of ways. The
standard approach uses values from the Highway Capacity Manual [\citet
{manual2010highway}] that provides a look-up table for road capacity
based on road type and number of lanes.

\begin{figure}[t]

\includegraphics{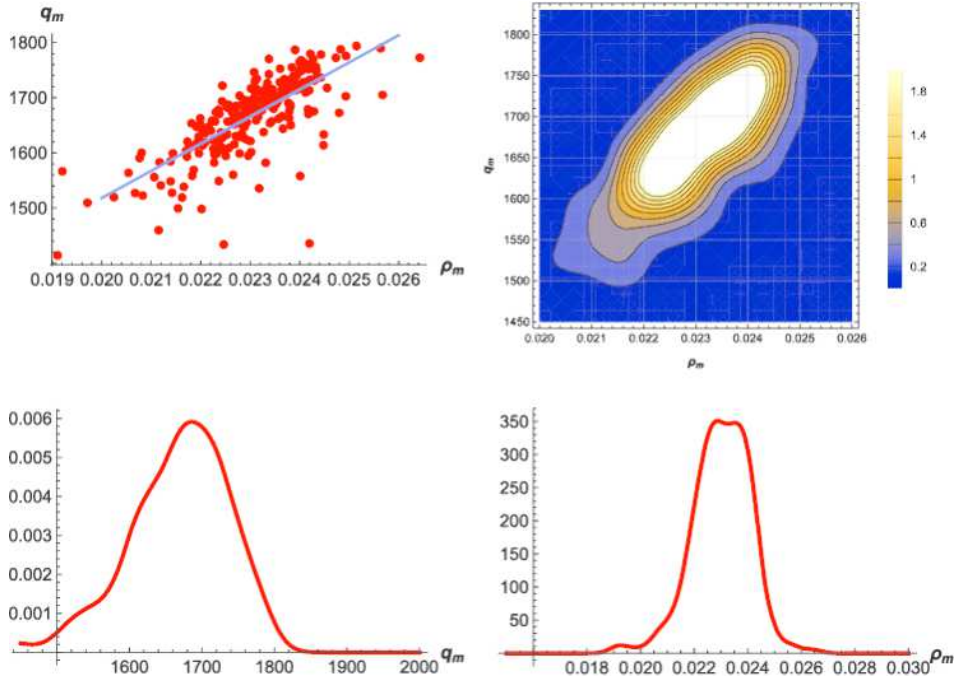}

\caption{Joint and marginal distributions for critical flow and
density.}\label{fig:phis}\label{fig:critical-densiti-flow}
\end{figure}

However, in practice, the parameters are not fixed and change over
time. To empirically illustrate the stochastic nature of the
parameters, we estimate capacity and critical density from the
measurements for 242 days in 2009, on a segment of interstate highway
I-55 in Chicago. Holidays and weekends as well as days with unreliable
measurements were excluded. Figure~\ref{fig:phis} plots $\rho_c$ and
$q_c$ across the days. Clearly, there is a linear relation between
$q_c$ and $\rho_c$. Road capacity can vary from day to day and its
distribution has a heavy left tail. On the other hand (for our data
set), the critical density $\rho_c$ has a relatively tight distribution
around the value 0.023~veh${}/{}$m.
%

This nonstatic nature of the parameters motivates the need for a
sequential online parameter learning algorithm.

\subsection{Traffic flow dynamics as a nonlinear state-space model}
Let $y_t$ denote the observed traffic density data and $ y^t=(y_1,
\ldots, y_t)$ be the current history of data. Let $\theta_t$ be a
hidden state vector of traffic densities. We assume that boundary
conditions $\rho_{0t}$ and $ \rho_{(M+1)t}$, which represent traffic
states on the downstream and upstream ends of a road segment, are
given, as well as an initial condition $\theta_0$. In practice,
boundary conditions are measure from sensors, such as loop detectors
and radars, available at both ends of a road segment, and initial
conditions either assume an empty road or state of traffic measured
from cameras or satellites.

We denote
\[
\theta_t = ( \rho_{1t}, \ldots, \rho_{Mt} ).
\]

The expectation conditional of the next state $E(\theta_{t+1}| \theta
_t) = f_\phi(\theta_t)$ is given by the solution of the LWR model. Here
$ \phi$ denotes unknown parameters. A numerical Godunov scheme computes
$ f_\phi(\theta_t)$ given the parameters, $\phi$, of the triangular
fundamental diagram. 

Our model has a state-space formulation of an observation and evolution
system given by
%
\begin{eqnarray}
\mbox{Observation:}&\quad& y_{t+1} = H_{t+1}
\theta_{t+1} + \varepsilon^v_{t+1};
\varepsilon^v_{t+1} \sim N(0,V_{t+1}),
\label{eqn-y}
\\
\mbox{Evolution:}&\quad& \theta_{t+1} = f_{\phi}(
\theta_t) + \varepsilon^w_{t+1};
\varepsilon^w_{t+1} \sim N(0,W_{t+1}),
\label{(eqn-x)}
\end{eqnarray}
where $V_{t}$ and $W_t$ are evolution and equation error, respectively,
and $y_{t+1} ={}$vector of measured traffic flow density,
$f_{\phi} ={}$LWR evolution equation calculated via Godunov's schema,
$\phi= (q_c,\rho_c, \rho_{\mathrm{jam}})$ triangular fundamental diagram
parameters.
The observation matrix $H_{t+1}$ picks out cells with measurements available.

\begin{figure}[t]

\includegraphics{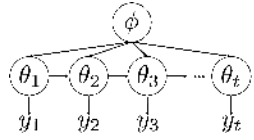}

\caption{Parameter learning graphical model.}
\label{fig:pl}
\end{figure}

Figure~\ref{fig:pl} provides the graphical model for the state
evolution structure. Our goal is to develop a particle filter to draw
samples from\vspace*{1pt} the filtered posteriors $p(\theta_t | y^t)$ and $p(\phi| y^t)$.
%
The operator $H_t: \mathbb{R}^{M} \rightarrow\mathbb{R}^k$ is the
measurement model that depends on the sensor type, and in our setting
we make it linear. In a simplest case $H_t = H$ is a projection
operator, which ``removes'' nonmeasured elements from the state vector.

%
\begin{figure}[b]

\includegraphics{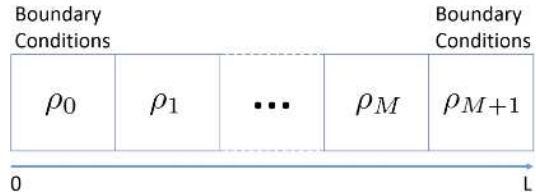}

\caption{Underlying state space for a road segment.}\label{fig:example}
\end{figure}

While we treat the parameters $\phi$ of the LWR model as static, our
model can easily be extended to allow for stochastic evolutions or
characteristics to govern the dynamics.

\subsection{Godunov's scheme}
The LWR model~(\ref{eqn:lwr1}) describes the evolution of traffic flow on
a road segment with uniform topology, as shown in Figure \ref{fig:example}
(see Appendix~\ref{app:model-derivation}). The change in road segment characteristics
(crossing, number of lanes, speed limit, curvature, etc.) can be
modeled using a junction. The treatment of junctions requires specific
efforts for physical consistency and mathematical compatibility with
the link model. For uniqueness of the solution of the junction problem,
different conditions have been used, for instance, maximizing the
incoming flow through the junction was suggested by \citet
{daganzo1995cell} and \citet{coclite2005traffic}. \citet
{holden1995mathematical} consider maximizing a concave function of the
incoming flow. A formulation using internal dynamics for the junction
is equivalent [\citet{lebacque2005first}] to the vertex models for the
merge and diverge junction; see \citet{garavello2006traffic} for more details.

Standard finite difference schemes are too inaccurate for solving the
LWR model; \citet{godunov1959difference} showed that a first order
finite difference scheme is inaccurate for calculating with a small
time step. Moreover, none of the second order difference schemes
preserve monotonicity of the $\rho_0$, and thus are not applicable.

Given an initial condition $\rho_0(x), x\in[0,L]$, propagating the LWR
model requires solving the associated Cauchy problem. If the initial
condition is piecewise constant (which is the case for many numerical
approximations) and self-similar, this reduces to a Riemann problem.
Godunov's scheme then solves a Riemann problem between each cell. This
is an initial value problem with initial conditions having a single
discontinuity
%
\begin{equation}
\label{eqn-initial} \rho_0(x) = \cases{ \rho_l, &\quad
$x<0$,
\cr
\rho_r, &\quad$ x>0$.}
\end{equation}
For the Riemann problem, the speed of the shock wave propagation is
given by the Rankine--Hugoniot relation (\ref{eqn:r-h}).
Heuristically, imagine at initial time $t=0$ that there are two regions
in the domain with different values of thermodynamic parameters (flow,
density and speed in our case). The two regions are divided by a thin
membrane and at the initial time the membrane is removed. The
computational problem is to find the values of thermodynamic parameters
at all future times.

According to Godunov's scheme, we calculate the iterates
%
\begin{equation}
\label{eqn-schema} \rho_{i}^{n+1} = \rho_i^n
+ \frac{\tau}{h} \bigl(q_G \bigl(\rho_{i-1}^{n},
\rho_{i}^{n} \bigr) - q_G \bigl(
\rho_{i}^{n},\rho_{i+1}^{n} \bigr)
\bigr),
\end{equation}
where $\rho_i^n$ is the density value at the point with coordinates $x
= ih$, $t = n\tau$, with $h$ a space discretization step and $\tau$ a
time discretization step.

The function $q_G(\rho_l,\rho_r)$ is defined by
%
\begin{equation}
\label{eqn-qg} q_G(\rho_l, \rho_r) =
\cases{ q(\rho_l), &\quad$ \rho_r < \rho_l
\le\rho_c$, \vspace*{1pt}
\cr
q(\rho_c), &\quad$
\rho_r \le\rho_c \le\rho_l$, \vspace*{1pt}
\cr
q(\rho_r), & \quad$ \rho_c \le\rho_r <
\rho_l$, \vspace*{1pt}
\cr
\min \bigl(q(\rho_l), q(
\rho_r) \bigr), &\quad$ \rho_l < \rho_r$.}
\end{equation}
Typically, a virtual cell is introduced on both sides of the domain to
include boundary conditions (in and out flow). This leads to a left boundary
\begin{eqnarray}
\rho_{0}^{n+1} = \rho_0^n +
\frac{\tau}{h} \bigl(q_G \bigl(\rho_{-1}^{n},
\rho_{0}^{n} \bigr) - q_G \bigl(
\rho_{0}^{n},\rho_{1}^{n} \bigr)
\bigr),
\nonumber
\\
\eqntext{\displaystyle\mbox{with } \rho_{-1}^n =
\frac{1}{\tau}\int_{(n-1/2)\tau}^{(n+1/2)\tau}\rho(0,t)\,dt,}
\end{eqnarray}
and right boundary
\begin{eqnarray}
\rho_{M}^{n+1} = \rho_0^n +
\frac{\tau}{h} \bigl(q_G \bigl(\rho_{M-1}^{n},
\rho_{M}^{n} \bigr) - q_G \bigl(
\rho_{M}^{n},\rho_{M+1}^{n} \bigr)
\bigr),
\nonumber
\\
\eqntext{\displaystyle\mbox{with } \rho_{M+1}^n =
\frac{1}{\tau}\int_{(n-1/2)\tau}^{(n+1/2)\tau}\rho(L,t)\,dt.}
\end{eqnarray}
Numerical stability in space and time is ensured by the
Courant--Friedrichs--Lewy type condition [\citet
{courant1928partiellen}]: $\tau\le h/|v_{\max}|$, where $v_{\max}$ is
the maximum wave velocity present in the meshed domain at any given
point in time.

\subsection{State uncertainty is a mixture distribution}\label{sec:mixture}
\begin{figure}

\includegraphics{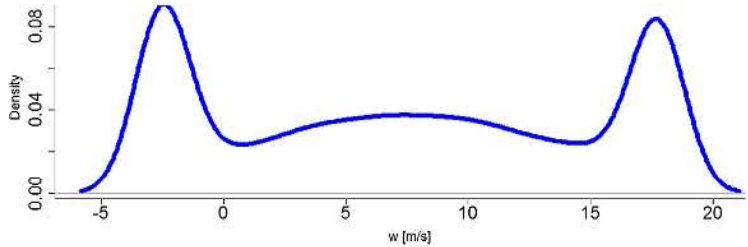}

\caption{Uncertainty of shock wave propagation speed.}
\label{fig:sigma-mixture}
\end{figure}

When uncertainty about the traffic state gets propagated from one time
step to another using Godunov's scheme, the current unimodal
distribution can update to a mixture distribution. For example, this
happens at the location of a shock wave, when the cell on the right is
in a free-flow regime and the cell on the left is in a congested
regime. This can be demonstrated by a simple Monte Carlo experiment.
Consider two consecutive cells, with densities $\rho_l$ and $\rho_r$
correspondingly, both following a truncated normal distribution. Assume
$\rho_l \sim TN(\mu= 0.02, \sigma= 0.01, a = 0,b=0.2)$ and $\rho_r
\sim TN(\mu=0.03, \sigma=0.01, a=0,b=0.2)$, where $a$ and $b$ are lower
and upper bounds of a truncated normal distribution correspondingly.
Using a triangular fundamental diagram with $q_c = 1600$ veh${}/{}$h, $\rho_c
= 0.025$~veh${}/{}$m, and $\rho_{\mathrm{jam}} = 0.2$~veh${}/{}$m, we can calculate the
speed of the shock wave propagation $w$ given by equation~\ref
{eqn:r-h}. We then simulate the distribution over $w$, using $N=1000$
samples. Figure~\ref{fig:sigma-mixture} shows the results of the
experiment. The uncertainty over speed propagation is a bimodal mixture
distribution, implying the uncertainty about the density at the future
times is also a mixture. Our example in Section~\ref{sec:numerical}
shows that the behavior of uncertainty about traffic flow density state
matches this bimodal shape found here.
%

\section{Particle filtering of the LWR model}\label{sec:analysis}

\subsection{A fully adapted particle filter}

Particle filtering methods are designed to provide sequential state
inference from the set of filtered posteriors $p(\theta_t | y^t)$;
see, for example, \citet{gordon1993novel,carpenter1999improved,pitt1999filtering,liu2001combined,storvik2002particle,car10}.
Our algorithm will be based on the \citet{liu2001combined} filter. The
major difference is a fully adapted filter that resamples first using
the predictive distribution and propagates forward using the
conditional posterior. The fully adapted filter mitigates particle
filter degeneracy, although the usual compounding of the Monte Carlo
errors still exists [\citet{godsill2004monte}].

The predictive likelihood for the next observation, $y_{t+1}$, is
required to implement our particle filter. Given the current state
variable $ \theta_t $, the predictive likelihood is defined by
\[
p ( y_{t+1}|\theta_t, \phi) =\int p ( y_{t+1}|
\theta_{t+1},\phi) p ( \theta_{t+1}|\theta_t,\phi)
\,d\theta_{t+1}.
\]
Propagation of states requires the conditional posterior for the next
state $p( \theta_{t+1} | \theta_t, \phi, y_{t+1}) $. This density can
be computed using the model assumptions via the system of distributions
\begin{eqnarray*}
p(y_{t+1}|\theta_{t+1},\phi) & \sim &N(H_{t+1}
\theta_{t+1}, V_{t+1}),
\\
p(\theta_{t+1}|\theta_t,\phi)& \sim &N
\bigl(f_{\phi}( \theta_t), W_{t+1} \bigr).
\end{eqnarray*}
Marginalizing out $\theta_{t+1}$ leads us to distributions
\begin{eqnarray*}
p(y_{t+1}| \theta_t, \phi) & \sim& N \bigl(H_{t+1}f_{\phi}(
\theta_t), H_{t+1}W_{t+1}H_{t+1}^T+V_{t+1}
\bigr).
\end{eqnarray*}
For propagation of $\theta_{t+1}$, we use Bayes' rule and the
conditional posterior
\[
p( \theta_{t+1} | \theta_t, \phi, y_{t+1} ) \sim N
(\mu_{t+1}, C_{t+1} ), %
\]
where the mean and variance $(\mu_{t+1}, C_{t+1})$ follow the Kalman
recursion [\citet{doucet2000sequential}]:
\begin{eqnarray*}
\mbox{Forecast:}&\quad& \mu_f = f_{\phi}(
\theta_t),\qquad C_f = W_{t+1},
\\
\mbox{Kalman Gain:}&\quad& K = C_fH_{t+1}^T
\bigl(H_{t+1}C_fH_{t+1}^T+V_{t+1}
\bigr)^{-1},
\\
\mbox{Measurement Assimilation:}&\quad& \mu_{t+1} = \mu_f
+ K(y_{t+1} - H_{t+1} \mu_f),
\\
&&{} C_{t+1} = (I - KH_{t+1})C_f.
\end{eqnarray*}

To develop our particle filter, we now factorize the joint conditional
distribution as
\[
p(y_{t+1},\theta_{t+1}|\theta_t,
\phi)=p(y_{t+1}|\theta_t,\phi) p(\theta_{t+1}|
\theta_t,\phi,y_{t+1}).
\]
The goal is to obtain the new filtering distribution $p(\theta
_{t+1}|y^{t+1})$ from the current $p(\theta_{t}|y^{t})$ and to provide
a particle approximation to the parameter posterior, $p(\phi|y^t)$. We
start with a particle (a.k.a. random histogram of draws) filtering
approximation to the joint distribution of the state and parameters,
denoted by
\[
p^N\bigl(\theta_t,\phi|y^{t}\bigr)=
\frac{1}{ N }\sum_{i=1}^N
\delta_{ (\theta_t,
\phi)^{(i)}}, %
\]
where $ \delta$ is a Dirac measure. As the number of particles
increases $N\rightarrow\infty$, the law of large numbers guarantees that
this distribution converges to the true filtered distribution $p(\theta
_{t},\phi|y^{t})$.

For the next marginal posterior distribution, the Bayes rule yields
\[
p^{N} \bigl( \theta_{t+1} |y^{t+1} \bigr) = \sum
_{i=1}^{N} w_{t}^{ ( i ) } p
\bigl( \theta_{t+1}| (\theta_t, \phi)^{ (i ) },y_{t+1}
\bigr), %
\]
where the particle weights are determined by
\[
w_{t}^{ ( i ) }=\frac{p ( y_{t+1}| ( \theta_t, \phi)^{ ( i )
} ) }{\sum_{i=1}^{N}p ( y_{t+1}| ( \theta_t, \phi)^{ ( i ) } )
}.
\]
The algorithm consists of three steps:
\begin{longlist}
\item[\textit{Step} 1.] (Resample) Draw an index $k(i) \sim \operatorname{Mult}_{N} (
w_{t}^{ ( 1 ) },\ldots,w_{t}^{ ( N ) } )$
for $i=1,\ldots,N$.

\item[\textit{Step} 2.] (Propagate) Draw $\theta_{t+1}^{ ( i ) }\sim p
( \theta_{t+1}| (\theta_t,\phi)^{k ( i ) },y_{t+1}
)$ for $i=1,\ldots,N$.

\item[\textit{Step} 3.] (Replenish) Draw $\phi^{(i)} \sim\frac{1}{N}
\sum_{i=1}^N \delta_{ [- \varepsilon, \varepsilon]} (\phi^{ k(i) } )$,
\end{longlist}
where\vspace*{2pt} $ \delta_{ [- \varepsilon, \varepsilon]} ( \cdot) $ denotes the
Dirac measure in an interval $[-\varepsilon, \varepsilon]$. Thus, we
resample $\phi^{k(i)}$ from mixture uniform distribution with support
$[\phi^{k(i)}-\varepsilon,\phi^{k(i)}+\varepsilon]$, $i=1,\ldots,N$,
and equal mixing rights. The jittering parameter $\varepsilon$ is used
to calculate unique $\phi^{i}$ particles.
Both $\theta^{(i)}_{t+1}$ in step~2 and $\phi^{(i)}$ in step~3 are
drawn based on resampled $\phi^{k(i)}$, thus the resampling creates a
new set of particles $ ( \theta_t, \phi)^{k ( i ) } $. Steps 1 and~2
of the algorithm were suggested in the auxiliary particle filter of
\citet{pitt1999filtering}.

It has been previously shown that particle filters suffer the
degeneracy issue when the number of particles is not sufficient [\citet
{bengtsson2008curse,snyder2011particle}]. However, our approach relies
on predictive likelihood and is less prone to a degeneracy issue, which
plagues standard sample-importance resample filters.

\section{Real-time accident modeling}\label{sec:study-capacity}
We illustrate our methodology on a data set from an accident on I-55.
We show how quickly our approach can identify a drop in capacity
(critical flow) due to an accident. On May 9, 2014, a semi-tractor
trailer caught fire (\citeauthor{accident-report}) at 6:40~AM on
interstate highway I-55 near Weber Road in Romeoville, Illinois, which
is a southwest suburb of Chicago. The police shut down the southbound
lanes. As is commonplace, the accident was visible from the other side
of the road, and the ``rubbernecking'' effect, drivers slowing down to
watch an accident, caused a dramatic reduction in capacity and congestion.

There are several reasons for capacity reduction during an accident.
Under normal conditions, an average time delay before a vehicle starts
accelerating following a leader is half a second. However, during an
accident there is a large difference in times that drivers took to look
at the accident location before accelerating. These results were
obtained by \citet{knoop2008capacity} via analyzed video taken by
helicopter from accident locations. Most of the vehicles would
accelerate at the usual rate out of the jam, and the shock wave would
move backward. However, it only requires a small fraction of drivers
that keep driving slowly until they reach the accident location to
cause large escape times at the location of the incident and hence for
the low capacity. There is also heterogeneity in acceleration delays
between left lane (closest to accident) and right lane. On average, in
the left lane, cars take longer before accelerating.

\begin{figure}[t]

\includegraphics{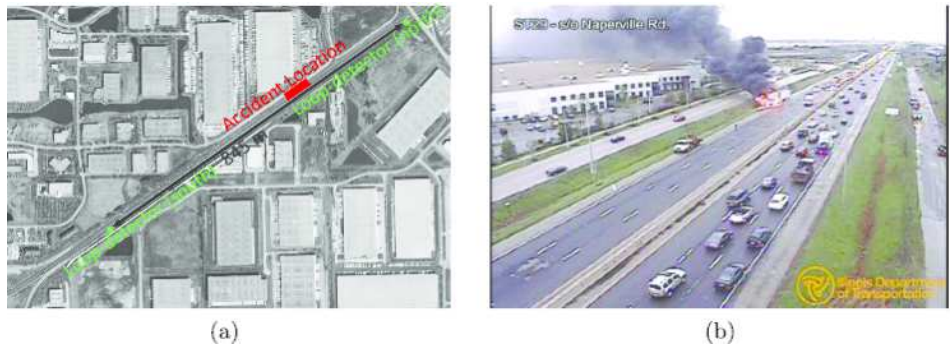}

\caption{Accident location.
\textup{(a)}~Accident and loop detector locations.
\textup{(b)}~Image of the accident from the roadside camera.
The left panel \textup{(a)} shows the satellite
image with the location of the accident identified (red rectangle) and
two loop detectors located (green circles) on the opposite direction
before and after the accident location (credit: Bing Maps). The right
panel \textup{(b)} shows the image of the truck on fire taken by Illinois
Department of Transportation's roadside camera on the day of the accident.}
\label{fig:accident-location}
\end{figure}

Figure~\ref{fig:accident-location}(a) shows the map location of the
accident. Figure~\ref{fig:accident-location}(b) shows two of the loop
detectors located before and after the accident location from which the
data was collected.
%

The length of the road segment between two loop detectors is 845 meters
and we discretized it with four cells, with each space step $h = 845/4
= 211$ meters and used time step $\tau= 5 $ minutes. This combination
of time and space step satisfies the Courant--Friedrichs--Lewy
condition so that numerical stability in space and time is ensured.
Further, a five minute interval was chosen since it is a standard
interval over which the measured data is averaged to provide smooth
input data. Our initial prior on road capacity is assumed to be
uniform, with $q_m \sim U[1440, 1560]$~veh${}/{}$h, and we set critical
density to $\rho_m = 0.025$~veh${}/{}$m; both are based on empirical
observations of typical ranges for those parameters as shown in
Figure~\ref{fig:critical-densiti-flow}. To replenish the parameters
(step 3 of the algorithm), we used $\varepsilon_{q_m} = 50$~veh${}/{}$h for
capacity and $\varepsilon_{\rho_m} = 0$~veh${}/{}$m for critical density, as
there is no learning for this parameter. The value for $\varepsilon
_{q_m}$ is based on empirical observations, that capacity change
usually does not exceed 50 veh${}/{}$h within a five minute interval.

We have chosen the measurement noise's standard deviation to be $0.2
\times10^{-2}$ veh${}/{}$m, and standard deviation for the evolution
equation error to be equal to $0.1 \times10^{-2}$ veh${}/{}$m. Given that we
did not have access to manufacturer's specifications of the loop
detectors, we use a value within the guidelines of the specification.
The error for the evolution equation was chosen to be consistent with
the results reported in~\citet{chu2011validation}, where authors report
that standard deviation of the LWR model forecast error is usually
under 3\%, but higher for congested flow when compared with observation
data from motorway sensors. In our numerical example, we use 4\%.

To address the problem of model identification, we utilize the relation
between free-flow speed, capacity and critical density, namely, $ v_f =
q_c/\rho_c$. Based on the data measured on a typical day during
off-peak hours, the free-flow speed $v_f \approx17$~m${}/{}$s. Our particle
weights are regularized by
\[
w_{t}^{ ( i ) }=\frac{p ( y_{t+1}| ( \theta_t, \phi)^{ ( i )
} ) \varphi(q_c^{(i)}/\rho_c^{(i)}, v_f, \sigma_{v_f}) }{\sum_{i=1}^{N}
[p ( y_{t+1}| ( \theta_t, \phi)^{ ( i ) } )\varphi(q_c^{(i)}/\rho
_c^{(i)}, v_f, \sigma_{v_f}) ]},
\]
where $\varphi$ is the p.d.f. of the normally distributed variable. The
prior error standard deviation was set at $\sigma_{v_f} = 5$~m${}/{}$s.
Choice of both $v_f$ and $\sigma_{v_f}$ is based on empirical observations.

\begin{figure}[b]

\includegraphics{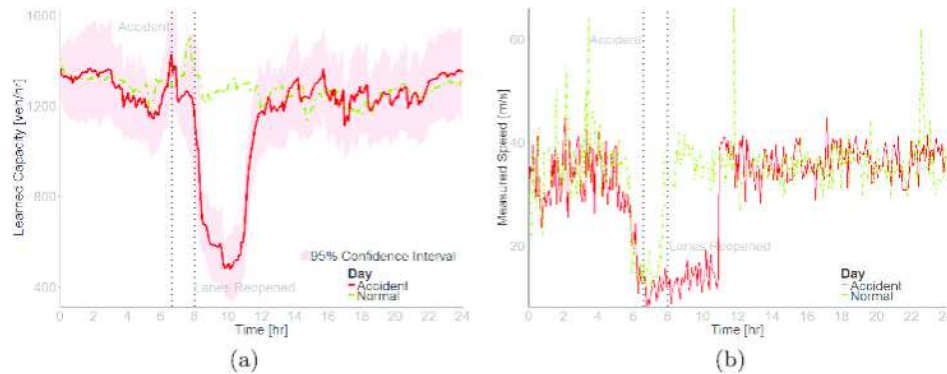}

\caption{Comparison of the learned capacity and measured speed on
Thursday, May 8th (normal day) and Friday, May 9th (accident day).
\textup{(a)}~Learned capacity of the road segment.
\textup{(b)} Measured speed at the south loop detector.
On both plots the left vertical line identifies the time when accident
happened (6:40 AM) and the second vertical lime corresponds to the time
when all of the lanes reopened at 8:00 AM, according to the news
report. The number of particles was chosen $N = 5000$. Accident data
was plotted using a solid line and normal day with a dashed line.}
\label{fig:accident}
\end{figure}

Figure~\ref{fig:accident}(a) compares the road capacity learned by the
algorithm on the day of the accident and from the previous day, which
was accident free, with similar weather conditions. Figure~\ref
{fig:accident}(b) shows the measured speed by the south loop detector
on both days. There is a time lag of approximately 15 minutes between
traffic flow speed reverts to a normal level and capacity recovers.
This time lag corresponds to three measurements (data is reported every
five minutes) and is explained by the time it takes the algorithm to
learn the capacity.
%

Our algorithm captures the effect of capacity degradation as a result
of the accident. We provide 95\% Bayes credible intervals to
demonstrate that uncertainty about the estimate is larger during the
normal operating mode and lower during the periods of capacity
degradation and recovery. If we compare the speed plot and capacity
plot in Figure~\ref{fig:accident}(a) and (b), we can see that the
slope of the speed curve on an accident day is much steeper than the
slope of the learned capacity curve; it is due to the fact that there
is some delay associated with the learning process. In other words, the
algorithm does not learn that the flow regime has recovered
instantaneously, but rather it takes three to five measurements before
it learns.

Under normal conditions, a full-width freeway lane has a capacity of
2000 passenger vehicles per hour [\citet{manual2010highway}], with a
truck being counted as 1.5 passenger cars. In Illinois, the loop
detectors give reliable data for the vehicle counts but not for
different vehicle classes and it is hard to identify the share of
trucks in the traffic flow; consequently, the flow is measured in
vehicles per hour and not in passenger car units per hour. Thus, the
learned capacity is around 1500 veh${}/{}$h on a normal day, that is
consistent with the theoretical estimate from the highway capacity
manual. On the accident day we detect a reduction in capacity of up to
66\%. This is similar to the results of \citet{knoop2008capacity} who
use helicopter images from Netherlands roads to observe a 50\%
reduction of capacity, due to the reduction of the discharge rate at
the bottleneck (accident location) due to rubbernecking. A larger drop
in our case might be explained by regional differences in driving
style. American drivers might be driving more carefully in the presence
of an accident, and by the fact that a truck on fire is more
``spectacular'' than a regular vehicle crash, with people spending more
time to observe. Such a drop in the flow rates is remarkable given the
absence of any physical obstacles.

\section{Calibration experiment}\label{sec:numerical}
The previous example illustrates a drop in capacity due to an accident.
However, we do not know if the drop size is properly estimated since
the ground truth is unobserved. To demonstrate that our algorithm
properly captures state and parameter dynamics, we must use simulated
data with a realistic traffic flow pattern. We simulate data that
mimics traffic flow on Chicago's I-55 highway. Figure~\ref
{fig:week-pattern} below shows traffic patterns on February 6th, 2009
(Friday) and all five work days of the following week (week of February 9th).
Several conclusions can be drawn from the traffic patterns:
\begin{enumerate}[(iii)]
\item[(i)] Break down start times are different from day to day, even on the
same day of the week (Fridays) of different weeks.
\item[(ii)] The duration of the flow at the lowest speed is different, with
Wednesday being the worst and Thursday the best.
\item[(iii)] The breakdown period is shorter than the recovery period.
\end{enumerate}
The latter point follows from the asymmetric shape of the triangular
fundamental diagram, where the free-flow speed $v_f$ (speed at which
drivers arrive to the end of the congestion queue) is higher than the
backward wave propagation speed $w$ (speed at which drivers depart from
the front of the congestion queue).

%
\begin{figure}[t]

\includegraphics{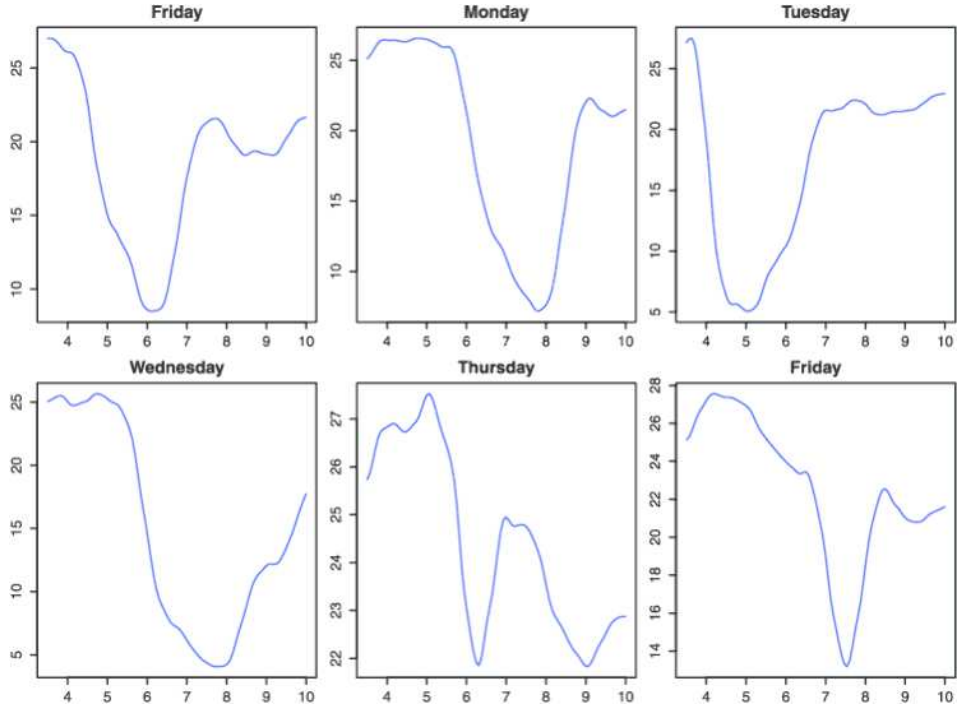}

\caption{Chicago I-55 Work Day Morning Peak Traffic Patterns. The
$x$-axis corresponds to the time of the day.}
\label{fig:week-pattern}
\end{figure}

\begin{figure}

\includegraphics{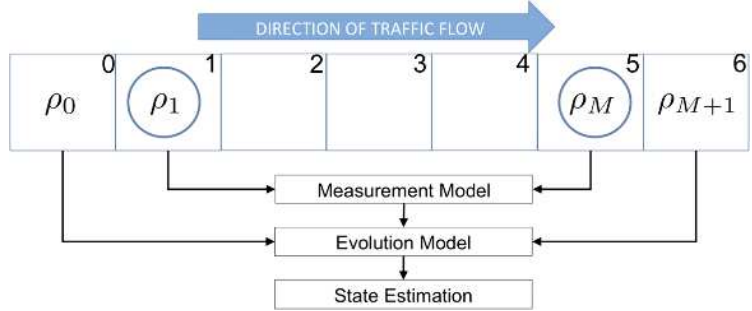}

\caption{Simulated stretch of a freeway. The arrow shows the direction
of the traffic flow. It is assumed that measurements from circled first
and last cells of the domain are available.}\label{fig:example1}
\end{figure}

Our road segment is 1.5 kilometers long and we choose a time horizon of
1600 seconds.
Figure~\ref{fig:example1} shows our road segment model and its
discretization scheme with five internal cells and 2 boundary cells.
Our discretization grid cell is of length $h=300$ meters and time
interval of $\tau= 5$ seconds. The initial conditions are set to be
uniform traffic density of 0.01~veh${}/{}$m.

%

\begin{figure}[b]

\includegraphics{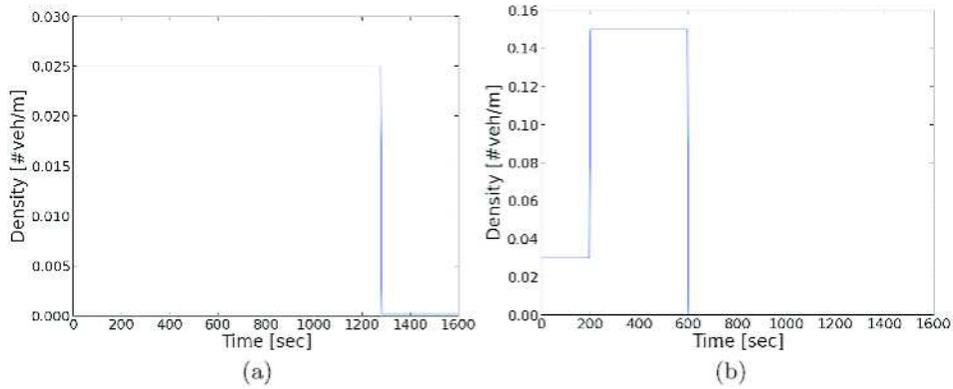}

\caption{Boundary conditions used to produce simulated data.
\textup{(a)} Left (cell 0).
\textup{(b)} Right (cell 6).}\label{fig:boundary}
\end{figure}

\begin{figure}[b]

\includegraphics{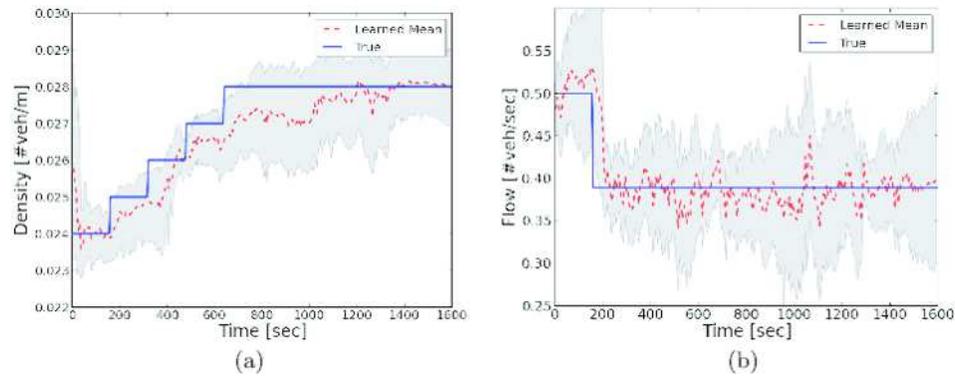}

\caption{True and learned values of parameters and 90\% confidence
interval band.
\textup{(a)}~Critical density learning.
\textup{(b)}~Capacity learning.}\label{fig:parameter-learning}
\end{figure}

To mimic a typical morning commute pattern, we have chosen boundary
conditions so that our simulated data set begins with a free-flow
traffic regime followed by a breakdown and then recovery. The breakdown
starts 3 minutes into the simulation and the recovery starts at the 10
minute mark. In the left boundary cell (cell 0) we have a constant
vehicle density followed by a drop in density to zero; see Figure~\ref
{fig:boundary}(a). It mimics the constant inflow of morning commuters
that eventually stops. On the right boundary cell (cell 6) we have
uncontested density at the beginning, followed by density of 0.145
veh${}/{}$m, which represents heavily congested traffic flow, and followed by
a drop in density to zero; see Figure~\ref{fig:boundary}(b). The right
boundary condition corresponds to a location where an on-ramp merges
into a highway. When the flow on the ramp is high, a bottleneck is
created at the merge location. The boundary conditions are shown in
Figure~\ref{fig:boundary}. Over the course of simulation we changed the
traffic flow parameters. Capacity and critical density parameters used
to produce simulated data are shown in Figure~\ref{fig:parameter-learning}.
%

To simulate the measured data, we compute the solution of the LWR model
for the measurement cells 1 and 5 and add noise to it. We have chosen
the noise standard deviation to be $0.8 \times10^{-2}$ veh${}/{}$m, and
standard deviation for the evolution equation error to be equal to $0.1
\times10^{-2}$ veh${}/{}$m.

\begin{figure}

\includegraphics{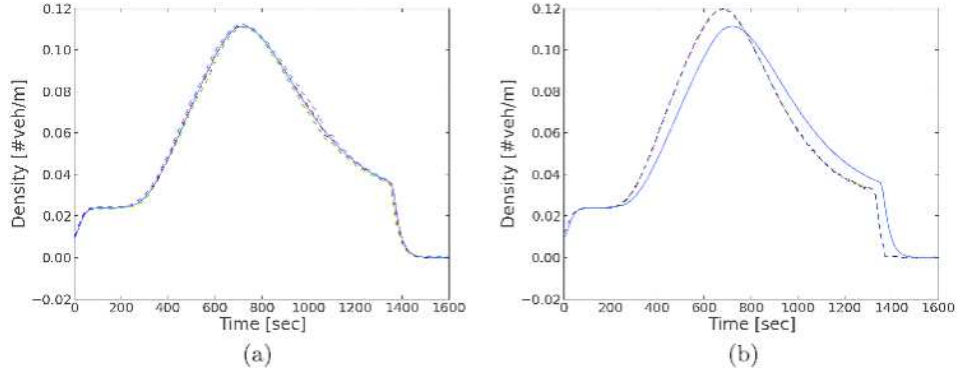}

\caption{Estimated density at cell 3.
\textup{(a)} With parameter learning.
\textup{(b)} Without parameter learning. The blue line on both plots is
ground truth and the dotted line is the filtered value computed by the
algorithm. The number of particles was chosen $N = 1000$.}\label
{fig:density-cell-3}
\end{figure}

Figure~\ref{fig:density-cell-3} compares the estimated traffic density
in cell 3 with the true simulated traffic density for two different
scenarios. In the first scenario we used the parameter learning step of
the algorithm and in the second scenario we kept capacity and critical
density parameters fixed. We can see the sensitivity to parameter
learning. Without learning, the density profile is shifted in a
meaningful way. Clearly, full Bayesian parameter learning corrects this bias.
%

%

To illustrate the dynamics of parameter learning, we change the LWR
parameters $\rho_c$ and $q_c$ several times throughout the simulation,
as mentioned above. In principle, we could also directly model $\phi_t$
with its own state evolution.

Figure~\ref{fig:parameter-learning} shows the expected value and
95th percentile of the filtered posterior distribution of
the model parameters. We can see, as expected, there is a certain delay
between the underlying parameter change and the filtering algorithm
captures~the change. Change in capacity is picked up faster than change
in critical density.

%
\begin{figure}

\includegraphics{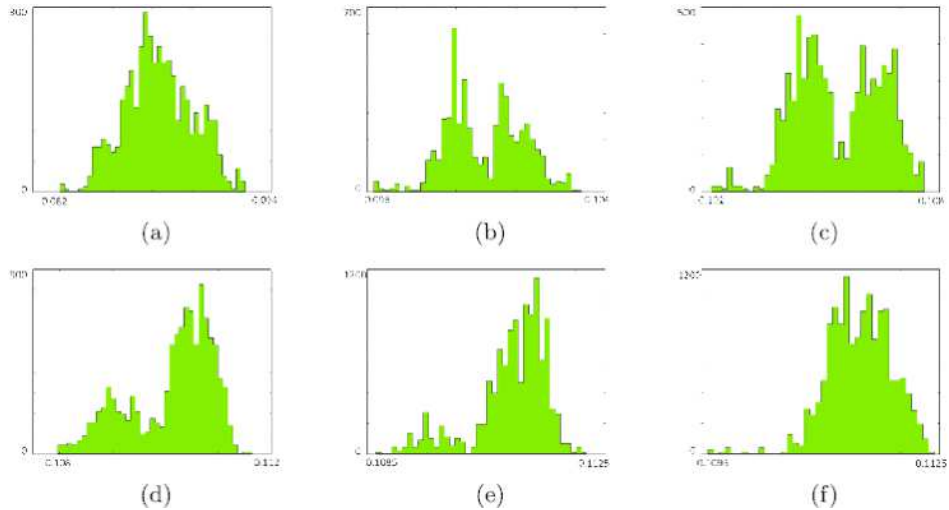}

\caption{Uncertainty distribution about traffic flow density estimated
by filtering algorithm at cell 3. Panel \textup{(a)} shows distribution at the
time right before the shock wave reaches the cell. Panels \textup{(b)--(e)}
show the time when the shock wave travels through the cell. Panel \textup{(f)}
shows distribution at the time after the shock wave moves beyond the
cell.
\textup{(a)}$~t=565$,
\textup{(b)}$~t=615$, \textup{(c)}$~t=645$,
\textup{(d)}$~t=680$, \textup{(e)}$~t=690$, \textup{(f)}$~t=700$.}\label{fig:mixture}
\end{figure}

In Section~\ref{sec:mixture} we showed that the distribution over
traffic density is a mixture distribution at the locations when the
density in the left cell is below critical density and the density in
the right cell is above. To further demonstrate this fact, Figure~\ref
{fig:mixture} shows the distribution over density at cell 3 before and
after the shock wave travels through the cell.

Figure~\ref{fig:density-cell-3} shows that a shock wave travels through
cell 3 between $t=600$ and $t=700$. We can see that the distribution
over state is unimodal at those time steps. However, in between, it is
a mixture.

Appendix~\ref{sec:kalman-recursion} develops the projection operator
$H$ and the necessary Kalman recursions for this and the following examples.

\section{Discussion}\label{sec:conclusion}
In this paper we analyze the LWR traffic flow model with application to
Chicago's interstate I-55 highway. We show how particle filtering and
learning provides a real-time estimate of the density states. We
sequentially learn the parameters of the fundamental diagram which is
the central input for the LWR dynamics of traffic flow. Our results
have a number of important implications for transportation system
management applications. In particular, a real-time assessment of model
states and parameters corrects the biases in estimating the current
density of states used for forecasting.

Our methodology quickly handles the drop in capacity due to a major
traffic accident on Chicago's interstate I-55 highway. We also use a
calibration study to show how close a filtered state vector is to the
true one. When measurements are sparse in space and the parameters are
fixed, pure filtering would misestimate the current state. However, our
approach corrects this by incorporating parameters~learning
simultaneously. This leads to an accurate estimation of traffic density.

There are a number of possible avenues for a future approach. First,
the LWR model is only valid when the relationship between flow and
density is time independent. Second, the model does not describe
traffic behavior within a queue or when particular instabilities such
as stop and go traffic exist. Third, the model is not realistic for
free-flowing traffic, as vehicle bypassing that happens frequently in
this regime is not captured. Although, from a system management
perspective, free-flow traffic is not an issue, extending our approach
to higher order traffic flow models will lead to improvements in estimation.

Developing methods to incorporate model monitoring is an important area
of future research. For example, alternative models might correspond to
different assumptions about the shape of the fundamental diagram. We
can statistically discriminate two models using a sequential likelihood
ratio (Bayes factor), $B_t$, given~by
\[
B_{t} = \frac{p(y_1,\ldots,y_t|M_1)}{p(y_1,\ldots,y_t|M_0)},
\]
where $p(y_1,\ldots,y_t|M_i) = \prod_{j=1}^{t} p(y_j|y_{1:j-1},M_i)$.
This is simply a product of marginal predictive densities, which the
particle filter approximates by
\[
p^N(y_j|y_{1:j-1},M) = \frac{1}{N}\sum
_{i=1}^{N}p \bigl(y_j |
\theta_{j-1}^i,M \bigr).
\]
Another avenue is to extend our particle algorithm to a transportation
network with simultaneous tracking of multiple segments. This will make
our methodology applicable to real-life transportation networks of a
large metropolitan area. Within our framework, it is feasible to filter
over the boundary conditions. This also applies in the case of GPS
probes, where inferring the boundary conditions is a hard task, since
location and time of the measurement are random and one rarely observes
the boundary conditions.

\begin{appendix}
\section{Derivation of flow model}\label{app:model-derivation}
Let $q(x,t)$, $\rho(x,t)$ and $v(x,t)$ denote traffic flow, density and
speed at position $x$ at time $t$.
Kinematic wave theory establishes a relationship between density $\rho$
and flow $q$, which is known as the \emph{fundamental diagram} given by
the functional equation $q(x,t)=q(\rho(x,t))$ where $q(x,t)$ is flow.
The conservation law implies that with no inflow or outflow
%
\begin{equation}
\label{eqn:conservative-law} \frac{\partial\rho(x,t)}{\partial t} + \frac{\partial q(x,t)}{\partial x} = 0.
\end{equation}

Combined with fundamental diagram function, we obtain the equation for
$\rho(x,t)$:
\[
\frac{\partial\rho(x,t)}{\partial t} + \frac{\partial q(\rho)}{\partial
\rho} \frac{\partial\rho(x,t)}{\partial x}= 0.
\]
The term $w = \partial q(\rho)/\partial\rho$ is called the wave
velocity. To get a more intuitive understanding of the problem, it is
convenient to use the cumulative flow $N(x,t)$, the number of vehicles
that pass location $x$ by time $t$. Then the conservation law can be
derived by evaluating
\[
\frac{\partial N}{\partial t} = q(x,t),\qquad \frac{\partial N}{\partial x} = -\rho(x,t).
\]
Assuming that $N(x,t)$ is smooth,
\[
\frac{\partial^2 N}{\partial x\,\partial t} = \frac{\partial^2
N}{\partial t\, \partial x},
\]
we get the conservation law (\ref{eqn:conservative-law}). In practice,
the function $N$ has discontinuity of the first kind (first derivative),
however, the conservation law holds in the case of discontinuities as
long as $N(x,t)$ is continuous along the shock path. The method of
characteristics can be used to solve the equation (\ref
{eqn:conservative-law}). Specifically, from (\ref
{eqn:conservative-law}) $\rho(x,t)$ is constant ($d\rho/ds = 0$) along
a characteristic curve (wave) described by
\[
\frac{dt}{ds} = q'(\rho).
\]
Eliminating $s$ gives
\[
\rho(x,t) = \rho \bigl(x - q'(\rho_0)t \bigr).
\]
Thus, density is constant along the straight line with slope $dq/d\rho$
(characteristic line) and the slope is nothing but a shock propagation
speed. For a free-flow speed the shock moves forward and for jammed
traffic it moves backward. In Newell's case the forward shock
propagation speed is $v_f$
and the backward shock propagations speed is given by $w$.

\section{Derivation of Kalman recursion}\label{sec:kalman-recursion}
Measurements are taken at the first and last cells of the road segment
and noise is independently distributed, with covariance structure
$ V_t = V = v I_2 $ and $ W_t = W= w I_5 $.
The operator $H_t$ and the Kalman gain matrix $K_t$ are of the
following form:
\[
H_t=H = \pmatrix{ 1 & 0 & 0 & 0 & 0
\cr
0 & 0 & 0 & 0 & 1 }\quad\mbox{and}\quad K_t = K =\pmatrix{ \displaystyle\frac{w}{v+w} & 0
\vspace*{3pt}\cr
0 & 0
\cr
0 & 0
\cr
0 & 0
\vspace*{3pt}\cr
0 & \displaystyle\frac{w}{v+w} }.
\]
This leads us to the following Kalman updates:
\begin{eqnarray*}
C_{t+1} &=& \pmatrix{\displaystyle w \biggl(1-\frac{w}{v+w} \biggr) & 0 & 0 & 0 &
0
\vspace*{3pt}\cr
0 & w & 0 & 0 & 0
\cr
0 & 0 & w & 0 & 0
\cr
0 & 0 & 0 & w & 0
\vspace*{3pt}\cr
0 & 0 & 0 & 0 & w \displaystyle \biggl(1-\frac{w}{v+w} \biggr) },
\\[3pt]
\mu_{t+1} &=& \biggl(\mu^f_1+
\frac{w (y_1-\mu^f_1 )}{v+w},\mu^f_2,\mu^f_3,
\mu^f_4,\mu^f_5+
\frac{w (y_2-\mu^f_5 )}{v+w} \biggr)^T.
\end{eqnarray*}
The variance of the predictive likelihood distribution is given by
\[
\mathit{HWH}^T + V = \pmatrix{ v+w & 0
\cr
0 & v+w }.
\]
\end{appendix}



%

\printaddresses
\end{document}